# University-industry research collaboration: a model to assess university capability[1]

*Giovanni Abramo*[a,b,*], *Ciriaco Andrea D'Angelo*[a], *Flavia Di Costa*[a]

[a] Laboratory for Studies of Research and Technology Transfer
School of Engineering, Department of Management
University of Rome "Tor Vergata"

[b] National Research Council of Italy

**Abstract**

Scholars and policy makers recognize that collaboration between industry and the public research institutions is a necessity for innovation and national economic development. This work presents an econometric model which expresses the university capability for collaboration with industry as a function of size, location and research quality. The field of observation is made of the census of 2001-2003 scientific articles in the hard sciences, co-authored by universities and private enterprises located in Italy. The analysis shows that research quality of universities has an impact higher than geographic distance on the capability for collaborating with industry. The model proposed and the measures that descend from it are suited for use at various levels of administration, to assist in realizing the "third role" of universities: the contribution to socio-economic development through public to private technology transfer.

## 1. Introduction

In recent years, the industrially-developed nations have identified cooperation between industry and public research institutions as a major policy priority (OECD, 2007). At the same time, scholars in the economics of innovation have analyzed the modalities through which public-private interaction develops. Beginning in the 1980s, there has been an intensification of studies that examine support systems for technology transfer, enhancement and exploitation of public research outputs (Bozeman, 2000). These first concentrated on the concept of the national innovation system (Freeman, 1987; Lundvall, 1992; Nelson, 1993) and in the 1990s turned to examining the regional innovation system (Cooke et al, 2004) and the "triple helix" model of national development (Etzkowitz and Leydesdorff, 1998), which presented close interaction between spheres of public research, industry and government institutions as the ideal method for increased innovation and regional development.

Public-private research collaboration is one of the main modes of technology transfer. A number of studies have investigated this phenomenon. In general this studies investigate the università-industry relations along two main dimensions: either the contribution of universities to the innovative activities of industry; or the ways in which the relations generate and actualize (D'Este and Patel, 2007; Muscio, 2010).

A line of investigation has concentrated on possible personal and institutional motivations, and has viewed collaboration as an exchange relationship that provides benefits to both partners (Meyer-Krahmer and Schmoch, 1998; Manjarrés-Henrıquez et al., 2009). Our study follows along this line of research. It starts with identifying environmental and intrinsic factors that favour development of public-private research collaboration to finally providing quantified measurement of their relative weight. In accordance with the existing literature, we will focus on three factors which determine the capability of collaboration of any particular university with private enterprises: i) the size of the university; ii) its geographic location; iii) the scientific excellence of the university researchers.

Enterprises that are initiating a search for collaboration view larger universities as offering larger research groups and a greater mix of disciplines. Meanwhile, research also shows that it is larger- rather than smaller-sized enterprises that have the tendency to be active in collaboration (Segarra-Blasco and Arauzo-Carod, 2008; OECD, 2007; Fontana et al., 2006). Studies that adopt a "gravitational model" (Ponds et al., 2007) also show that the capability for collaboration between pairs of public research institutions and private enterprises from different systems (meaning, for example, from different territories) depends on the product of their respective masses, and also on the square of the distance between the pair. Lee and Mansfield (1996) back up the conclusion that even if the reputation of a university carries weight, geographic distance is a determining factor, probably because of associated costs. Lindelof and Lofsten (2004) also showed that for new technology-based firms, nearness to a university favors information and idea exchanges through formal and informal networks. Meanwhile, excellence in universities in a particular region has positive consequences for innovation, which even extend to neighboring regions (Jaffe, 1989; Jaffe et al., 1993), while the flow of knowledge from the public sector to industry weakens progressively with increasing distance (Arundel and Geuna, 2004). Generally, the number of collaborations between partner pairs decreases exponentially with increase in distance between the two partners (Katz, 1994). There has also been ample study of the geographic aspect of "knowledge spillovers". Scholars first measured this phenomenon through "coincidence" or "concentration" index (Jaffe, 1989, Audretsch and Feldman, 1996), then through citation of patents (Jaffe et al., 1993), and most recently using spatial econometrics (Anselin, 1998; Anselin et al., 2000; Fingleton and López-Baso, 2006). In particolar, when dealing with knowledge spillovers one one should distinguish Marshall-Arrow-Romer (MAR) spillovers from Jacobs spillovers (Carlino, 2001; Glaeser et al., 1992). The MAR externalities concern knowledge spillovers among firms in an industry. The concentration of firms of the same sector in a geographical area facilitates knowledge transfer among them and thus, innovation and economic growth. Porter (1990) too states that knowledge spillovers in specialized, geographically concentrated industries are able to foster economic growth and that local competition facilitates rapid adoption of innovation. On the contrary, according to the Jacobs spillover theory, factors that foster innovation and growth are the variety and diversity of geographically proximate industries rather than geographical specialization.

Geographic proximity seems especially important for partnership between organizations from different institutional backgrounds (Ponds et al., 2007), although it is also useful to examine other dimensions of proximity, such as cognitive, social and organizational similarity (Boschma, 2005).

Besides the distance factor, the quality of research supplied by a university can also play a role in determining intensity of scientific collaborations, specifically the presence of researchers with a high level of scientific productivity and impact. Various studies have shown correlation between universities' excellence and their intensity of collaboration with private enterprises (Balconi and Laboranti, 2006; Van Looy et al., 2004, Barnes et al., 2002). Abramo et al. (2009a), examining the Italian situation, demonstrated that academic researchers who collaborate with the private sector have research performance that is superior to the researchers who tend not to enter into collaborations. However, further analysis showed that enterprises do not necessarily choose partners with the best scientific performance (Abramo et al., 2009b).

In this study we propose to quantify the weight of factors noted in the literature in determining universities' capability to establish research collaborations with private enterprises. The study examines the Italian system, specifically collaborations between universities and enterprises situated within the country. Findings need to be interpreted within the historical, cultural and structural characteristics of the context. Differently, from most Oecd countries, R&D spending in the public sector in Italy is about the same as in the private sector. To this anomaly, can be added at least another two of a structural nature, rendering the process of university-industry collaboration, on the one hand, undoubtedly more difficult, and, on the other hand, even more necessary. The first is the progressive demise of the Italian high-tech industry, as shown by the worsening specialization indexes of the Italian economy and the revealing historical analysis of Gallino (2003)[2]. The second is the composition of the Italian industrial system, characterized by a disproportionate ratio of small and micro companies, and a concentration of private research spending in northern Italy. The effects deriving from the synergistic interaction of these three anomalies are aggravated by the cultural humus of poor interaction and collaboration amongst public research institutions and industry. The recently sought reversal in this tendency encounters difficulty in taking off due, to the structural distance that has been created between the two systems.

---

[2]Gallino describes how flourishing high-tech sectors of the Italian industry, such as the chemical, computer, airplane, electronic industries either disappeared or irremediably shrunk in the past 40 years.

In this work we study university-industry collaborations which have led to the co-joint publication of articles in international scientific journals during the 2001-2003 triennium. Although bibliometric approaches do present certain limits, co-authored publications remain as one of the most tangible and best documented indicators of collaboration in the hard sciences (Melin and Persson, 1996; Katz and Martin, 1997; Laudel, 2002). Given the paucity of academic patents in Italy (Abramo, 2007), such publications remain as the most representative indicator available in Italy. Also, compared to other empirical methods based on partial surveys, bibliometric analyses present clear advantages in terms of scope and field of observation.

Data on collaborations observed through co-authorship will be analyzed and aggregated at the level of single universities. The number of collaborations between a specific university and private enterprises will be viewed as a dependent variable, and an econometric model will be used to analyze the factors that could influence such capability of collaboration: i) the "mass", or the size of the research group able to supply collaboration to private enterprise; ii) the geographic distance between the university and the private enterprises demanding for research collaboration; iii) the scientific excellence of the researchers within the scientific field in which collaboration would take place. The study will be organized in two stages or levels. First, the analysis will be conducted at the micro-level of individual scientific fields: the econometric model will be used to examine, for all universities, the capability of collaboration by the research groups active in each scientific field. Secondly, the study proceeds to an analysis by university, aggregating the data for all the research fields active in the university.

To the authors' knowledge no previous works studied all three factors together. Consequently, no study so far has weighted against one other the relative importance of each of the factors, so that tradeoffs between them could be assessed. Moreover, the bibliometric approach utilized, differently from the traditional surveys, made it possible to base the investigation on a very large number of observations, which made the analysis quite robust.

The following section of this paper lays out the organization of the research, the methodologies adopted, the data sources and field of observation for the study. Section 3 identifies the main characteristics of research collaborations in Italy. Section 4 presents the econometric model used and the results of its application at the level of single scientific field and at the aggregate level. The last section presents conclusions and the final considerations of the authors.

## 2. Research methodology and organization

The first step of this work is to identify the determinants of university capability for collaboration with industry, which are both significant and measurable. Many determinants are possible, including social and institutional factors, but it is not always possible to obtain reliable and complete data or to identify representative proxies. The factors most suited to quantitative analysis have been chosen for study. The study develops two econometric models in which capability for collaboration of universities with industry is expressed through sets of explanatory variables. The first model considers only quantitative structural variables, meaning variables independent of the merit of universities. These quantitative variables are sources of competitive advantage in establishing research collaboration with industry: i) the size of the university and ii) its location with respect to geographic distribution of enterprises that express a demand for collaboration. The second model integrates these structural variables with others that represent the scientific quality of the university and its individual research groups. This second set of variables represents the distinctive competencies of the university, which are also sources of competitive advantage. The working definition of the second group of variables is methodologically more delicate. The two models can be used to provide separate estimation of collaboration capability of universities due to i) economic rents not linked to the universities' merit and ii) the merit or excellence of the university in research. When assessing and comparing the performance of universities in collaboration, the effect of structural variables independent of merit should be identified and taken into account.

The two structural factors considered by the first model must certainly play a role in selection for collaboration. A higher number of researchers definitely provides a greater supply of competencies, while geographical proximity offers greater possibilities for formal and informal contact and reduces the costs of collaboration. It is more complex and delicate to quantify scientific quality and the prestige of universities and their internal research groups. Italy does not have a well-established practice of evaluating public research, including research by universities, so it may not be easy for enterprises to find thorough information on competencies and comparative performance of research groups or single scientists. This means that the choice of a university partner by an enterprise often develops not only on the basis of objective information on quality but also through personal contacts. Choice is conditioned by "social proximity"[3]. This complex phenomenon cannot be modeled by a single variable, however part of the phenomenon is included in the variables of the models, since it is reasonable to expect positive correlation between social proximity and the university's size, geographical proximity and scientific excellence. Finally, different research and innovation policies and initiatives at regional level to foster university-industry collaborations may have an additional impact on the phenomenon under study.

In this paper, the study of university capability to collaborate with industry is conducted at two distinct levels: at the level of scientific field (from now on called scientific disciplinary sector, SDS, in accordance with the Italian university classification) and at the aggregate level. Each SDS has different characteristics in terms of the supply of collaboration by universities (size, location, quality) and the demand observed by industry (size and location). Analysis at the level of single SDS is essential if distortions typical of aggregate analysis are to be avoided. The two levels of analysis require different means of calculation for the explanatory variables: for the aggregate level of analysis it is necessary that the values for the variables be aggregated from those observed for the individual SDS, through normalization that takes account of the different frequencies of collaborations realized in the respective SDS.

The analysis was conducted using a negative binomial econometric model[4], which estimates the capacity of each variable to explain the capability for a university to collaborate with industry and to quantify its relative influence.

---

[3] The term "social proximity" is used in the sense given by Boschma (2005): "Social proximity is defined here in terms of embedded relations between agents at the micro-level. Relations between actors are socially embedded when they involve trust based on friendship, kinship and experience".

[4] This model was selected because the dependent variable is a count variable.

**2.1 Data source and field of observation**

As a source of data on university-enterprise collaboration, the authors referred to co-authored articles in international scientific journals, produced by academic and industry-based researchers. A number of bibliometricians recommend caution in the use of co-authorship based indicators as a source of evidence for true scientific collaboration. It has been noted that some forms of collaboration do not generate co-authored articles and some co-authored articles do not reflect actual collaboration (Katz and Martin, 1997). However analysis of co-authorship has indeed become one of the standard ways of measuring research collaborations, evidently because it does offer notable advantages that counterbalance its limitations (Melin and Persson, 1996; Laudel, 2002; Tijssen, 2004; Lundberg et al., 2006). Moreover the publications indicator is quantifiable and its measurement is non-invasive. Finally, in this specific study, the bibliometric census observed as proxy (more than 1,500 publications, representing a total of almost 2000 university-industry collaborations in the 2001-2003 triennium), guarantees a level of significance that can not be reached through sample-based surveys.

An analysis of patent applications filed by both private enterprises and universities could provide a useful complement to the publications indicator, but in Italy the number of patents filed by universities is notoriously low. Using the Espacenet search engine we have identified 284 patents filed by Italian universities between 2001 and 2003[5]. Of these, only 20 were co-filed with private enterprises. Because co-filing does not necessarily imply co-authorship[6], the number of university patents stemming from research collaboration with industry may be still less than 20. Meanwhile, many university researchers who file patents would also wish to publish scientific articles on the subject, therefore the analysis of publications could capture information concerning patents filed. For these reasons we have chosen not to include patents in our analysis. Finally, in comparison to empirical methods based on partial surveys, bibliometric analysis has the advantage of a large and thorough field of observation.

The specific data source for this study is the *Observatory of Public Research* (ORP), a bibliometric database developed by the authors. The database consists of all publications by all Italian research institutions indexed by Thomson Reuters Web of Science (WoS).

The census of university-enterprise co-authorship requires identification and standardization of all possible variations in the names given for Italian universities and enterprises situated in Italy. Using a complex algorithm for "disambiguation", each publication with at least one address for an Italian university is identified with its respective academic authors (for details see Abramo et al., 2008). In Italy, each university researcher must necessarily declare an affiliation with a specific SDS, and through this fact it is possible to link each publication and each collaboration to the SDS of the university author. There are actually 370 officially recognized disciplinary sectors in Italy[7] but this study considers only publications from scientists in the 183 SDSs that make up the 8 "hard science" disciplines: mathematics and computer sciences; physics; chemistry; earth sciences; biology; medicine; agricultural and veterinary sciences;

---

[5] After the introduction of "academic privilege" in 2001, university researchers are probably filing a higher number of patents, however identification and collection of relevant data is extremely difficult.
[6] Enterprises may have simply funded the academic research underlying co-filed patents.
[7] The complete list is available at http://www.miur.it/atti/2000/alladm001004_01.htm

industrial and information engineering.

There were 78 Italian universities in existence during the 2001-2003 triennium and 68 of these had at least one researcher active in one of the SDS under study, for a total of around 30,000 researchers. Collaborations were seen in a total of 141 out of the 183 hard sciences SDSs. There were 483 private enterprises involved in co-authorship of the scientific publications resulting from these collaborations.

**2.2 Definition of research collaboration**

In this study "research collaboration" indicates a collaboration between two or more organizations (at least one university and one private enterprise) that has resulted in a co-authored scientific publication. Collaborations can be studied both at the level of the organizations involved (university-enterprise) and at the level of the scientific sector involved (SDS-enterprise).

By university-enterprise collaboration we mean a research collaboration between a university and a private enterprise that has resulted in exactly one co-authored publication in the dataset under consideration. A publication by $m$ universities and $n$ private enterprises corresponds therefore to $m*n$ university-enterprise collaborations. The 1,534 publications seen in the dataset correspond to 1,983 separate university-enterprise collaborations and there were 1,226 university-enterprise pairs involved in these collaborations[8].

The study of collaborations at the SDS-enterprise level first required identification of the scientific disciplinary sector of the university co-author[9]. Given a publication co-authored with a private enterprise, the number of SDS-enterprise collaborations then represented is equal to the number of SDSs to which the university co-authors adhere. Again, the 1,534 publications seen in the dataset correspond to 2,363 SDS-university collaborations, with 1,775 distinct SDS-enterprise pairs involved in these collaborations. The difference in numbers of university-enterprise and SDS-enterprise collaborations (1,534 compared to 2,363) results from the fact that each publication represents one or more collaborations at an organizational level (universities-enterprises), which may potentially involve a different number of sectors (SDS-enterprises). Table 1 sums up the calculation of collaborations seen in the 1,534 publications in the dataset.

| Level of analysis | N. of collaborations | N. of pairs |
|---|---|---|
| *University-enterprise* | 1,983 | 1,226 |
| *SDS-enterprise* | 2,363 | 1,755 |

*Table 1: University-enterprise and SDS-enterprise pairs and collaborations in the dataset*

---

[8] The number of university-enterprise pairs is less than the number of university-enterprise collaborations because, in the period considered, some pairs collaborated more than once.

[9] Each SDS can be considered as a community of university researchers with a common set of competencies even if each SDS includes multiple and probably non-quantifiable specializations that continuously evolve. The authors maintain that the approximation introduced by considering the SDS as a unique set of competencies shared by the participating researchers permits the understanding of certain aspects of university-industry collaboration.

## 3. University-industry research collaborations in Italy

Sorting the research collaborations for the 2001-2003 triennium by single university (Table 2) shows that of the 68 universities with research personnel in the hard sciences SDSs, 10 do not show any collaborations with the private sector. Half of the universities together produce only 13% of the total collaborations, while three universities (Bologna, Milan and Padua) produce 20% of the total, with over 100 collaborations each. These three universities represent 15% of the total research staff employed in Italian universities, for the 141 hard sciences SDSs under study. Five very small universities (Scuola Superiore Sant'Anna of Pisa, University of Reggio Calabria "Mediterranean", University of Benevento "Sannio", University of Teramo, University of Viterbo "Tuscia") bring up the rear for number of collaborations per university. This table thus illustrates a strong correlation between number of collaborations and size of the university, measured as research personnel employed in the 141 SDSs actually observed as involved in collaborations (coefficient of correlation 0.83).

| University | Coll. | University | Coll. |
|---|---|---|---|
| University of Bologna (1582) | 153 | University of Trento (131) | 20 |
| University of Milan (1454) | 129 | University of Salerno (239) | 19 |
| University of Padua (1310) | 116 | Polytechnic University of Ancona (307) | 18 |
| University of Rome "La Sapienza" (2532) | 98 | Second University of Naples (610) | 18 |
| University of Pisa (1123) | 88 | University of Cagliari (602) | 18 |
| University of Pavia (690) | 78 | University of Udine (317) | 18 |
| University of Turin (1046) | 77 | University of Lecce (202) | 17 |
| Polytechnic of Milan (547) | 67 | San Raffaele Vita - Salute University – Milan (39) | 14 |
| University of Naples "Federico II" (1660) | 65 | University of Camerino (204) | 14 |
| University of Catania (953) | 65 | University of Eastern Piedmont "A. Avogadro" (151) | 14 |
| University of Florence (1087) | 65 | University of Venice "Ca' Foscari" (124) | 9 |
| University of Genova (1000) | 59 | University of Bergamo (28) | 8 |
| University of Perugia (703) | 59 | University of Calabria (234) | 8 |
| University of Siena (472) | 59 | University of Varese "Insubria" (193) | 7 |
| University of Ferrara (451) | 49 | University of Cassino (81) | 7 |
| University of Modena and Reggio Emilia (520) | 49 | University of Sassari (400) | 7 |
| University of Parma (707) | 47 | International School for Advanced Studies of Trieste (47) | 6 |
| University of Rome "Tor Vergata" (750) | 46 | University of "Roma Tre" (181) | 6 |
| University of Verona (304) | 45 | University of Basilicata (182) | 6 |
| Polytechnic of Turin (470) | 43 | University of Urbino "Carlo Bo" (139) | 5 |
| University of Milan "Bicocca" (279) | 37 | University of Catanzaro "Magna Grecia" (112) | 4 |
| Sacred Heart Catholic University (790) | 36 | University of Foggia (82) | 4 |
| University of Trieste (506) | 36 | Polytechnic of Bari (157) | 3 |
| University of Messina (867) | 30 | University of Rome Bio-medical Campus (38) | 2 |
| University of Chieti "Gabriele D'Annunzio" (251) | 28 | University of Reggio Calabria "Mediterranean" (94) | 1 |
| University of L'Aquila (392) | 28 | University of Benevento "Sannio" (69) | 1 |
| University of Brescia (249) | 27 | University of Viterbo "Tuscia" (131) | 1 |
| University of Bari (925) | 24 | Scuola Superiore Sant'Anna of Pisa (20) | 1 |
| University of Palermo (1000) | 23 | University of Teramo (46) | 1 |

*Table 2: Number of collaborations between Italian universities and private enterprises; in brackets, average number of scientists in the 141 SDSs under observation for 2001-2003 period*

Only 2 of the top 10 universities for number of collaborations are located in southern Italy (Naples and Catania), while 3 are located in central Italy (Rome "La Sapienza", Pisa, Florence); the other 6 are located in the north. Aggregating the data by region we can get a better view of the correlation between university location and intensity of university collaboration with the private sector (Table 3): more than half of the total of collaborations (55.9%) involve universities from northern Italy, with universities from central Italy following (26.5% of total collaborations) and the south having the smallest share (15.8% of collaborations).

These empirical results show with certainty that: i) the universities most active in collaborations with private enterprises are those situated in the northern Italy, which is historically more industrially developed than the remainder of the nation; ii) number of collaborations is strongly correlated to size of the university. The results support the hypothesis that size and geographic location of a university are the first determinants in its ability to establish collaborations with private research partners.

| Region | Number of collaborations by universities, per region | Frequency (%) |
|---|---|---|
| Lombardy | 367 | 18.5 |
| Emilia Romagna | 298 | 15.0 |
| Veneto | 170 | 8.6 |
| Piedmont | 134 | 6.8 |
| Friuli Venezia Giulia | 60 | 3.0 |
| Liguria | 59 | 3.0 |
| Trentino Alto Adige | 20 | 1.0 |
| *Sub Tot. North* | *1,108* | *55.9* |
| Tuscany | 213 | 10.7 |
| Lazio | 160 | 8.1 |
| Umbria | 59 | 3.0 |
| Abruzzo | 57 | 2.9 |
| Marche | 37 | 1.9 |
| *Sub Tot. Center* | *526* | *26.5* |
| Sicily | 118 | 6.0 |
| Campania | 103 | 5.2 |
| Apulia | 48 | 2.4 |
| Sardinia | 25 | 1.3 |
| Calabria | 13 | 0.7 |
| Basilicata | 6 | 0.3 |
| *Sub Tot. South* | *313* | *15.8* |
| Other* | 36 | 1.8 |
| *Total* | *1,983* | |

*Table 3: Number of collaborations by universities of each region (2001-2003)*
*\* The Sacred Heart Catholic University has locations in more than one region. The universities of two other regions (Molise and Valle d'Aosta) did not register any collaborations in 2001-2003.*

## 4. The econometric analysis

The census of data on university-enterprise collaborations with classification by SDS permits several levels of aggregation in the quantitative analysis of the effect of specific variables on the capability for a university to activate collaborations. Two distinct econometric models were used in order to separately distinguish and estimate the effects on collaboration capability from: i) structural variables independent of merit (i.e. size and location) and ii) variables indicating the scientific excellence of the university and its researchers.

The dependent variable in our study is the capability for a university to activate collaborations with industry, measured by the number of collaborations by each university with enterprises. Model 1 considers two independent variables: the size of the university and its geographical distance with respect to the enterprises that activated collaborations with universities in the 2001-2003 triennium. Model 2 includes the size and location variables, as well as two variables related to the research quality of the university for the SDSs that participate in collaboration.

The analysis was conducted first at the level of the single SDS and then through aggregation, at the level of the university as a whole. The working definitions of the single variables are slightly different for the single SDS and the aggregate analysis.

Formally the two econometric models can be described by:

$$CP_{SDS} = f(m_{SDS}, d_{SDS}) \qquad [1]$$
$$CP_{SDS} = f(m_{SDS}, d_{SDS}, excell_{SDS}, star\_scientist_{SDS}) \qquad [2]$$

where $f()$ denotes the regression function.

Here we present the single SDS analysis (the aggregate analysis is described in Section 4.2), where for a university and a specific SDS of interest:

$CP_{SDS}$ = capability for collaboration;
$m_{SDS}$ = number of research staff;
$d_{SDS}$ = distance from the enterprises that activated scientific collaborations, calculated as the weighted sum of the distances between the university and the capital cities of the administrative regions where enterprises are located[10]. The weighting depends, for each capital, on the number of collaborations between the specific SDS and the enterprises in that region, relative to the total of the SDS's collaborations with enterprises throughout Italy;
$excell_{SDS}$ = measure of average quality of the SDS research staff;
$star\_scientist_{SDS}$ = concentration of star scientists.

For the analysis of the university as a whole the data from each SDS are aggregated, with weighting for each SDS according to its size relative to the total census of national collaborations (for details, see the opening paragraphs of Section 4.2).

Measurement of the variables indicative of scientific quality ($excell_{SDS}$; and $star\_scientist_{SDS}$) relies on the ORP. The ORP allows calculation of the bibliometric performance and comparative level of quality of individual scientists, research groups (for example the SDS of a given university) and entire research institutions (for example a university). Here the quality indicator defined as Scientific Strength (SS) is used, where the SS of a single scientist is a count of the articles he or she has published in the period 2001 to 2003, with each one weighted according to the prestige of the publishing journal (percentile rank of impact factor, IF)[11]. "Star scientist" is defined as a researcher that places in the top 10% of ranking for Scientific Strength in the SDS to which he or she belongs[12]. Concentration of star scientists, then, is the ratio of two ratios: the numerator being the ratio of the number of star scientists in the SDS under consideration of university $i$ to the total number of star scientists in the same SDS; the denominator being the ratio of number of scientists in the SDS of university $i$ to the total number of scientists of all universities in the same SDS.

---

[10] Expressed as the straight-line surface distance between university and the capital of each administrative region.

[11] To account for the variability in IF seen in the journals of the different WoS subject categories, the IF value of each journal is expressed as the percentile ranking of all the journals in the same discipline. For example, a value of 90 indicates that 90% of the journals falling in the same discipline have lower impact factor than the one under consideration. The authors are aware of the limitations of this use of the Impact Factor as proxy for the quality of a publication (Moed and Van Leeuwen, 1996; Weingart, 2005), but believe that the scope of the present study justifies its use in absence of data on the number of citations received by each article.

[12] For details see Abramo et al., 2009c.

The SS of an SDS for a given university was measured on the basis of the whole of the publications produced by the researchers belonging to that particular SDS. This value was normalized to the number of the researchers for that SDS in the specific university. The result is a measure of quality weighted productivity ($excell_{SDS}$) that defines the excellence of the SDS of each specific university relative to those of all other Italian universities.

The two variables identified as indicative of quality (average quality of the entire SDS: $excell_{SDS}$; and concentration of star scientists in the SDS: $star\_scientist_{SDS}$) reflect factors that could influence the selection of a university partner by an enterprise. The hypothesis is that the capability of a university to realize collaborations with industry increases with the scientific excellence of its scientists. This hypothesis is supported by a previously completed empirical analysis of the same dataset (Abramo et al., 2009a). This previous analysis showed that the researchers who collaborate with industry are ones that are generally more productive than others. The reliability of observations from these two indicators of scientific excellence, and the fact that they concern an entire population of universities, are strong points of the model of analysis proposed in this study. The authors recognize that the model omits variables that deal with social proximity, which has been previously indicated as a determinant in the choice of academic partners (Prabhu 1999; Abramo et al., 2008b). However the analysis of these determinants would require survey techniques and an inferential methodology, with resulting limits in thoroughness of data and accuracy of estimates[13] compared to the quantitative methodology proposed here.

Although in a somewhat indirect manner, the explicative variables used in the two models do include some information that refers to social proximity. For example, it is logical that where groups of academic researchers develop social relations with colleagues in industry, relationships will be stronger for the universities with greater numbers of scientists. Social relations also benefit from geographic proximity of the organizations and individuals, because of the informality of contacts that such proximity permits. Finally, scientific excellence of universities and individual researchers should bring about major exposure and visibility and by implication greater opportunities for social encounters.

**4.1 Analysis at the level of individual scientific disciplinary sectors**

The two models were applied for each of the 141 SDSs that had been involved in university-enterprise collaborations. As an example of the analysis at the level of single SDS, we can examine the case of the pharmacology SDS. This particular sector, along with electronics, internal medicine and biochemistry, are the four SDSs with the greatest number of university-enterprise research collaborations. Tables 4 and 5 present descriptive statistics and the correlation coefficients between the explicative variables for the 44 universities with scientists belonging to the pharmacology SDS.

In the 2001-2003 triennium the 44 universities active in the pharmacology sector realized 105 collaborations, an average of 2.4 per university. Fifteen universities had no collaborations with industry. On average, each research group in the pharmacology sector had 14.2 researchers, but there was wide variation. Four universities had only one scientist in the sector (University of Naples "Parthenope", International School for

---

[13] Such an analysis would also have significant costs in time and funds.

Advanced Studies of Trieste, University of Foggia, University of Benevento "Sannio"). The largest research group, with 69 scientists, was at the University of Milan.

| Variable | Definition | Obs | Mean | Std. Dev. | Min | Max |
|---|---|---|---|---|---|---|
| $CP_{SDS}$ | Capability for collaboration | 44 | 2.386 | 3.171 | 0 | 13 |
| $m_{SDS}$ | Number of scientists | 44 | 14.227 | 12.629 | 1 | 69 |
| $d_{SDS}$ | Distance (km) | 44 | 368.405 | 201.465 | 173.902 | 857.281 |
| $excell_{SDS}$ | Scientific impact per scientist | 44 | 266.345 | 262.248 | 0 | 1623.112 |
| $star\_scient_{SDS}$ | Concentration of star scientists | 44 | 0.933 | 1.606 | 0 | 9.429 |

*Table 4: Descriptive statistics for regression variables, pharmacology SDS*

| | $m_{SDS}$ | $d_{SDS}$ | $excell_{SDS}$ | $star\_scient_{SDS}$ |
|---|---|---|---|---|
| $m_{SDS}$ | 1.000 | | | |
| $d_{SDS}$ | -0.165 | 1.000 | | |
| $excell_{SDS}$ | -0.058 | -0.000 | 1.000 | |
| $star\_scient_{SDS}$ | 0.039 | 0.032 | 0.905*** | 1.000 |

*Table 5: Correlation among variables, pharmacology SDS*
*Statistical significance: *p-value <0.10, **p-value <0.05, ***p-value <0.01.*

The average weighted distance among universities and enterprises engaging in collaborations is seen as 368 km. The minimum distance was 174 km (University of Parma) and the maximum was 857 km (University of Catania).

Only two universities show a null value for the indicator *excell*$_{SDS}$, indicating a total absence of scientific publications by researchers in the pharmacology SDS. The maximum value (1623) is seen for the International School for Advanced Studies of Trieste, however this same university did not register any collaborations with the private sector. Finally, a full 20 universities (45%) do not have any star scientists in pharmacology, while the max value of star scientist concentration is recorded at International School for Advanced Studies of Trieste.

Table 5 presents correlations among variables. We note the absence of correlation among the independent variables in the model, with the exception of the pair of *star_scient*$_{SDS}$ and *excell*$_{SDS}$ (0.905). This particular correlation is expected, since a larger concentration of top scientists in a particular SDS raises the average quality of that SDS (the case of the International School for Advanced Studies of Trieste is a good example). The high value of VIF together with the strong correlation between the independent variables *star_scient*$_{SDS}$ and *excell*$_{SDS}$ induce to remove the former from the model.

The results from negative binomial regression applied to Models 1 and 2 are presented in Table 6. Incidence Rate Ratio (IRR)[14] results are shown in Table 7. The values of the likelihood ratio (LR) *test*[15] and *p-value* show that the response variable is over-dispersed, confirming the suitability of the negative binomial model over the

---

[14] Regression coefficients were interpreted as the difference between the log of expected counts, where formally, this can be written as β = log(μ$_{x0+1}$) - log( μ$_{x0}$ ), where β is the regression coefficient, μ is the expected count and the subscripts represent the predictor variable evaluated at x$_0$ and x$_0$+1 (implying a one unit change in the predictor variable x). The difference of two logs is equal to the log of their quotient, log( μ$_{x0+1}$ ) - log( μ$_{x0}$ )= log( μ$_{x0+1}$ / μ$_{x0}$ ), and therefore, we could have also interpreted the parameter estimate as the log of the ratio of expected counts. This explains the "ratio" in terms of incidence rate ratios (http://www.ats.ucla.edu/stat/stata/output/stata_nbreg_output.htm).

[15] The likelihood-ratio chi-square test indicates that the dispersion parameter alpha is equal to zero (H$_0$: *α=0;* no-overdispersion). The rejection of the null hypothesis would suggest that the dependent variable is over-dispersed and is not sufficiently described by the simpler Poisson distribution.

Poisson regression model.

For Model 1, the results show that the sign of the coefficient estimates for the independent variables $m_{SDS}$ and $d_{SDS}$ agrees with expectations, though only the coefficient of $m_{SDS}$ (number of researchers in the SDS being examined) is statistically significant (Table 6). The IRR values presented in Table 7 show that the estimated rate ratio would be expected to be 1.049 fold higher (+4.9%) for one-unit increases in the variable $m_{SDS}$, with other variables in the model held constant.

For Model 2, only the regression coefficient for $m_{SDS}$ and $d_{SDS}$ are statistically significant. Examination of the IRR values shows that with a one-unit increase in $m_{SDS}$ the estimated rate ratio increases by a factor of 1.051 times (+5.1%) while with a one-unit increase in distance $d_{SDS}$ the rate ratio decreases by a factor of 0.998 times (-0.2%).

|  | Model 1 | Model 2 |
| --- | --- | --- |
| $m_{SDS}$ | 0.048 (0.012)*** | 0.050 (0.014)*** |
| $d_{SDS}$ | -0.001 (0.001) | -0.002 (0.001)** |
| $excell_{SDS}$ |  | 0.002 (0.001) |
| const | 0.398 (0.453) | -0.081 (0.452) |
| n° obs | 44 | 44 |
| LR chi square α = 0 | 16.76*** | 15.06*** |

*Table 6: Negative binomial regression results predicting university-industry collaborations in pharmacology*
*Dependent variable: number of collaboration; method of estimation: negative binomial regression; robust standard errors shown in brackets.*
*Statistical significance: *p-value <0.10, **p-value <0.05, ***p-value <0.01.*

|  | Model 1 | Model 2 |
| --- | --- | --- |
| $m_{SDS}$ | 1.049 (0.013)*** | 1.051 (0.015)*** |
| $d_{SDS}$ | 0.999 (0.001) | 0.998 (0.001)** |
| $excell_{SDS}$ |  | 1.002 (0.001) |
| n° obs | 44 | 44 |
| LR chi square α = 0 | 16.76*** | 15.06*** |

*Table 7: Incidence rate ratios for the negative binomial regression predicting university-industry collaborations in pharmacology*
*Dependent variable: capability for collaboration; method of estimation: negative binomial regression; robust standard errors shown in brackets*
*Statistical significance: *p-value <0.10, **p-value <0.05, ***p-value <0.01.*

The coefficient of $d_{SDS}$, representing the distance between a university and industrial demand, is negative, revealing a proximity effect. Every 100 km increase in distance $d_{SDS}$ results in a 16% decrease in the capability number of collaborations. The University of Camerino and University of Ferrara provide an example of the actual calculation and effects of the $d_{SDS}$ variable. By road, these universities are actually 230 km apart, while the calculation of the weighted $d_{SDS}$ variable for the two universities gives a difference of approximately 100 km. Ignoring the effect of other variables (size and scientific excellence), estimates from Model 2 then indicate that the University of Camerino has a 16% lower capability for collaboration than the University of Ferrara, due to its less favorable geographic location with respect to the distribution of private enterprise demand for collaboration.

In general, it is difficult for the sectorial analysis to provide us with general indications of the determinants for university-enterprise collaboration. The results must be read in light of the distinctive sectorial characteristics that can influence the behavior of universities and enterprises in collaboration. There are specificities of working methods and organization of researchers for every scientific sector, and thus also sectorial differences in the management and organization of public-private research collaboration[16].

**4.2 Aggregate analysis**

For every university the data for all the SDSs in which it is active, was aggregated with weighting for each SDS according to its share of the total collaborations seen in the national census. The working definitions of the variables are different than those for the analysis at the SDS level. Here for the generic university:

$CP$ = capability for collaboration;
$m$ = weighted sum of researchers for all the SDSs, with weighting equal to the relative contribution of each SDS to the total of all enterprises collaborations at the national level;
$d$ = weighted distance from the enterprises that activated scientific collaborations; the weighting for each capital depending on the number of collaborations by enterprises in that region relative to the total of the university's collaborations at the national level calculated as the weighted sum of the distances between the university and the enterprises capital cities of the administrative regions where enterprises are located;
$star\_scientist$ = weighted sum of star scientists concentration, with weighting equal to the contribution of each SDS to the total of all enterprises collaborations at the national level.

The working definition for scientific excellence (*excell*) is more critical. The aggregation of the data relative to the SDS active in the university must take account not only of the incidence of each SDS on the national total of collaborations, but also of the different publishing "fertility" of the various SDSs. Thus we define the variable:

$$excell = \frac{1}{\sum_{SDS=1}^{N} m_{SDS}} \sum_{SDS=1}^{N} \frac{eccell_{SDS}}{PQ_{SDS}} \cdot m_{SDS} \cdot p_{SDS} = \frac{1}{\sum_{SDS=1}^{N} m_{SDS}} \sum_{SDS=1}^{N} \frac{SS_{SDS}}{PQ_{SDS}} p_{SDS}$$

where:
$SS_{SDS}$ = Scientific Strength for a specific SDS;
$PQ_{SDS}$ = National average of Scientific Strength for all universities active in the SDS;
$m_{SDS}$ = total researchers for a specific SDS.
$p_{SDS}$ = relative contribution of collaborations from a specific SDS to the national total;
$N$ = number of SDSs;

This formulation permits a rating for the average quality of each university that is not affected by the distortions that would occur in aggregate bibliometric analyses, which do not distinguish among scientific sectors (for details see Abramo et al., 2008).

Examining the correlation coefficients among the explicative variables indicates a very strong correlation between *star_scientist* and the variables *excell* and *m* (Table 8), signaling the possible presence of collinearity among the variables. A successive measure of VIF confirmed presence of collinearity, and as a result, the *star_scientist* variable was removed from the model.

---
[16] For example, the internal medicine disciplinary sector would certainly have different practices than the electronics or biochemistry SDS (these are the three SDSs showing highest frequency of collaboration with enterprises).

|  | m | d | excell | star_scientist |
|---|---|---|---|---|
| *m* | 1.000 | | | |
| *d* | -0.031 | 1.000 | | |
| *excell* | 0.629*** | -0.101 | 1.0000 | |
| *star_scientist* | 0.875*** | -0.217* | 0.712*** | 1.000 |

*Table 8: Correlation between variables*
*Statistical significance: \*p-value <0.10, \*\*p-value <0.05, \*\*\*p-value <0.01.*

Model 2 was utilized at aggregated level also, in to limit the problems of collinearity and obtain reliable estimates for the standard errors and confidence intervals:

$CP = f(m, d, excell)$ [Model 2]

The descriptive statistics for the Model 1 and Model 2 variables are presented in Table 9. The observations refer to all 68 universities with researchers appertaining to the SDS in the field of observation, including those where there are no recorded university-enterprise collaborations.

| Variable | Obs | Mean | Std. Dev. | Min | Max |
|---|---|---|---|---|---|
| *CP* | 68 | 29.162 | 33.985 | 0 | 153 |
| *m* | 68 | 6.179 | 7.071 | 0.003 | 36.076 |
| *d* | 68 | 378.615 | 178.475 | 207.211 | 837.826 |
| *excell* | 68 | 1.949 | 1.186 | 0 | 4.544 |
| *star_scientist* | 68 | 0.410 | 0.497 | 0 | 2.025 |

*Table 9: Descriptive statistics for regression variables*

The data should be read in the light of the working definitions for each variable, meaning that the data in Table 9 can not be directly compared to those for the pharmacology SDS (Table 4), since they result from aggregation of weighted data. For example, for *excell*, Table 9 shows a mean value (1.949) and a range (0, 4.544) that are completely different from those for *excell$_{SDS}$* in Table 4 (mean 266.345 and a range of 0 to 1623.112). This is a result of differences in the specific calculations for the "quality" variable at the individual and aggregate levels.

The results obtained from negative binomial regression applied to Model 1 and Model 2 are presented in Table 10. The results from IRR analysis are presented in Table 11. The values for *LR chi square α* confirm the suitability of the econometric model used.

The results obtained for Model 1 are consistent with expectations. Both of the regressors are statistically significant. The variable *m* (weighted sum of researchers for all the SDSs) is seen to have a positive influence on the capability of a university to collaborate with industry. The results also confirm the negative effect of distance. In Table 11 (results of IRR) we can see that a one-unit increase in the *m* variable and with the other variables of the model held constant is associated with IRR that increases by a factor of 1.175 (+17.5%). A one-unit increase in the distance variable *d* is associated with a decrease in the IRR by a factor of 0.999 (-0.1%).

|  | Model 1 | Model 2 |
|---|---|---|
| *m* | 0.161 (0.027)*** | 0.090 (0.023)*** |
| *d* | -0.001 (0.001)*** | -0.001(0.001)* |
| *excell* |  | 0.669 (0.173)*** |
| *const* | 2.439 (0.296)*** | 1.283 (0.555)** |
| *n° obs* | 68 | 68 |
| *LR chi square α = 0* | 704.78 *** | 319.42*** |

*Table 10: Negative binomial regression results predicting university-industry collaboration*
*Dependent variable: capability for collaboration; method of estimation: negative binomial regression; robust standard errors shown in between brackets*
*Statistical significance: \*p-value <0.10, \*\*p-value <0.05, \*\*\*p-value <0.01.*

|  | Model 1 | Model 2 |
|---|---|---|
| *m* | 1.175 (0.031)*** | 1.094 (0.026)*** |
| *d* | 0.999 (0.001)*** | 0.999 (0.001)* |
| *excell* |  | 1.952 (0.338)*** |
| *n° obs* | 68 | 68 |
| *LR chi square α = 0* | 704.78*** | 319.42*** |

*Table 11: Incidence rate ratios for the negative binomial regression predicting university-industry collaborations*
*Dependent variable: capability for collaboration; method of estimation: negative binomial regression; robust standard errors shown in brackets*
*Statistical significance: \*p-value <0.10, \*\*p-value <0.05, \*\*\*p-value <0.01.*

From the estimated coefficients for Model 2 we see that with a one-unit increase in the variable *m* the estimated rate ratio would be expected to increase by a factor 1.094 times higher (+9.4%, compared to +17.5% for Model 1). As for the Model 1, a one-unit increase in distance *d* reduces the rate ratio by a factor of 0.999 (-0.1%). For each 100 km increment in distance we would observe a 13.9% decrease in the number of collaborations in Model 1 and a 10.4% decrease in Model 2. These are solid indications of the occurrence of proximity effect. Finally, with a one-unit increase in the indicator *excell*, the rate ratio is 1.952 times higher (95.2%). The variable for indicator of average quality in Model 2 thus seems to be the one with the most notable effect: a one-unit increase in the indicator of excellence seems to result in twice the number of capability collaborations.

## 5. Conclusions

This paper presents a study of research collaboration between universities and private enterprises in Italy. The study identifies and analyses factors that influence the choice of university partners by enterprises. A bibliometric dataset is prepared by census of university-enterprise collaborations for the 2001-2003 triennium in Italy. An econometric model is then used to quantify the effect of several variables that jointly determine the capability for a university to establish collaborations with enterprises.

A university's capability for collaboration with the enterprises throughout Italy is expressed first as a function of two independent variables: the number of the university's researchers in the SDS being examined and the university's distance from the distribution of the demand from industry for collaboration. Next the authors augmented their model by adding two further variables relating to the scientific excellence of the university.

The addition to the knowledge in the literature up to this point consists in considering for the first time these factors together. This study has weighted against one other the relative importance of each of the factors, and tradeoffs between them has been assessed.

The analysis was first conducted at the level of single SDS and successively at the aggregate level of the entire university. The paper gives the example of the pharmacology sector to illustrate the analysis at the SDS level. The analysis of the pharmacology SDS in all universities gave results that were significant but difficult to generalize to the overall university-enterprise collaboration in Italy because of the characteristics of this particular scientific sector in Italy, especially the forms of interaction with private enterprises.

The aggregate level model seems more promising as a source of information that could be useful in planning policy interventions. However, from both levels of analysis, it can be observed that the capability of a university to attract private enterprise collaboration is influenced by the size of the group of academic researchers. This agrees with what we could expect, since larger universities will be better able to offer competencies in a large number of disciplines. The presence of a proximity effect is also confirmed, meaning that the capability of a university to collaborate decreases with increasing distance from enterprises. This effect is more notable for the pharmacology SDS than for universities as a whole.

Scientific excellence is evaluated with reference to a university's productivity and to the quality of its research output, and weighted to take account of the differing frequencies with which collaborations occur in each SDS. The excellence of the university is shown to be the most important determinant in explaining the capability for collaboration with enterprises. However, it is also reasonable to expect that even with disciplinary sectors that demonstrate a high level of excellence, a university could have a slim possibility of establishing partnerships if its "supply" is not in sectors that have adequate industrial demand, especially in relatively close geographic proximity.

The most notable difference between the two types of analysis is the difference in the effect of scientific excellence on the dependent variable. In analysis at the aggregate level the value of the estimated coefficient is very high, while in the analysis for the pharmacology SDS the coefficient is close to zero and not statistically significant[17]. The models presented in this paper provide insights into the complex phenomenon of university-enterprise collaboration which, though simplified, are clearly useful for practical purposes. The quantification of the effect of the independent variables in explaining the capability of universities could prove useful at various administrative levels. The authors recognize the limitations of the model proposed, especially for the limited number of variables considered and the choice of proxies for their measurement. However, the model presents significant new elements with respect to other possible evaluation systems of the capability for a university to activate collaborations. The model relies on objective data concerning the co-authorship of scientific articles. Implementation is possible by desk analysis, is non-invasive, relatively economical, and can be repeated over time. The analysis is not limited to sampling of specific sectors or organizations, but instead considers all the hard sciences disciplines for the totality of universities. The specific variable of "excellence" for research organizations is measured using a state of the art model. Finally, the authors are confident that it is possible to further strengthen this approach to render it more robust for future applications.

---

[17] A possible explanation is that private enterprises in pharmacology tend to draw heavily on networks of personal contacts in the choice of university partners for collaboration. Social proximity is indeed one of the primary determinants of choice in a research partnership but we have argued that the variables in the models embed some of the proximity considerations involved.

university interaction: Evidence from Spanish firms. *Research Policy*, 37(8), 1283-1295

Tijssen R.J.W., (2004). Is the commercialisation of scientific research affecting the production of public knowledge? Global trends in the output of corporate research articles. *Research Policy*, 33(5), 709-733.

Van Looy B., Ranga M., Callaert J., Debackere K., Zimmermann E., (2004). Combining entrepreneurial and scientific performance in academia: towards a compounded and reciprocal Matthew-effect? *Research Policy*, 33(3), 425-441.

Weingart P., (2005). Impact of bibliometrics upon the science system: Inadvertent consequences? *Scientometrics*, 62(1), 117-131.
20